\documentclass{PoS}
\usepackage{amsmath}
\usepackage{amssymb}
\usepackage{graphicx}
\usepackage{cite}
\usepackage{tabularx}
\title{Perturbing QCD with external fields}

\ShortTitle{Perturbing QCD with external fields}

\author{Paolo Cea \\ Universit\`a di Bari,   \& INFN - Bari, Italy\\
E-mail:  \email{paolo.cea@ba.infn.it}}

\author{\speaker{Leonardo Cosmai}\\
        INFN - Bari, Italy\\
        E-mail: \email{leonardo.cosmai@ba.infn.it}}

\author{Massimo D'Elia  \\  Universit\`a di Genova \& INFN - Genova, Italy\\
E-mail:  \email{massimo.delia@ge.infn.it} }

\abstract{We present some up-to-date results on QCD phase transition
in a chromomagnetic constant background field and in an abelian
monopole background field. Our results indicate that the QCD
critical temperature is not modified by a monopole background field,
whilst it is modified by a constant chromomagnetic field. We improve
our earlier estimate of the QCD critical parameters in a
chromomagnetic background field by performing lattice simulations
with weaker fields.}

\FullConference{The XXV International Symposium on Lattice Field Theory\\
         July 30 - August 4 2007\\
         Regensburg, Germany}

\begin{document}

\newcommand{\be}{\begin{equation}}
\newcommand{\ee}{\end{equation}}

%%%%%%%%%%%%%%%%%%%%%%%%%%%%%%%%%%%%%%%%%%%%%%%%%%%%%%%%%%%%%%%%%%%%%%%
\section{Introduction}
\label{Introduction}
In previous studies~\cite{Cea:2002wx,Cea:2005td}  on the vacuum
dynamics of pure non-abelian gauge theories we found that the
deconfinement temperature depends on the strength of an external
abelian chromomagnetic field.  We also verified that the same effect
is not present in the case of abelian gauge theories, suggesting
that it could be linked to the non-abelian nature of the gauge
group. In our opinion the dependence of the deconfinement
temperature on applied external fields is a consequence of the
dynamics underlying color confinement and therefore, apart from
possible phenomenological implications, such an effect could shed
light on confinement/deconfinement mechanisms.  On these premises it
is important to test if the effect continues to hold and how it
qualitatively changes when switching on fermionic degrees of freedom
and on that ground to investigate the dependence of the
deconfinement temperature on the strength of an external abelian
chromomagnetic field in the case of  full QCD with two flavors.
Besides another relevant issue regards the relation between
deconfinement and chiral symmetry restoration. As it is well known,
the two phenomena appear to be coincident in ordinary QCD, while
they are not so in different theories (like QCD with adjoint
fermions~\cite{Karsch:1998qj,Engels:2005rr,Lacagnina:2006sk}). Even
if  a simple explanation of this fact is still lacking it could be
strictly connected to the very dynamics of color confinement. A
contribution towards a clear understanding of this phenomenon could
be to study whether it is stable against the variation of external
parameters. In the present we investigated whether the deconfinement
temperature depends on the strength of an external abelian
chromomagnetic field and whether deconfinement and chiral symmetry
restoration for QCD with two flavors coincide even in presence of a
constant chromomagnetic field.

\section{A constant chromomagnetic field on the lattice}
\label{extfields}

QCD dynamics in presence of a time-independent background field at finite temperature
can be studied~\cite{Cea:1997ff,Cea:1999gn,Cea:2004ux} by means of the free energy functional
\be \label{freeenergy} {\mathcal{F}}[\vec{A}^{\text{ext}}] =
-\frac{1}{L_t} \ln \left\{
\frac{{\mathcal{Z}_T}[\vec{A}^{\text{ext}}]}{{\mathcal{Z}_T}[0]}
\right\} \; . \ee
with $\mathcal{Z}_T$ the thermal partition functional
\begin{eqnarray}
\label{ZetaT} \mathcal{Z}_T \left[ \vec{A}^{\text{ext}} \right]  &=
&\int_{U_k(L_t,\vec{x})=U_k(0,\vec{x})=U^{\text{ext}}_k(\vec{x})}
\mathcal{D}U \,  {\mathcal{D}} \psi  \, {\mathcal{D}} \bar{\psi}
e^{-(S_W+S_F)} \nonumber \\&=&
\int_{U_k(L_t,\vec{x})=U_k(0,\vec{x})=U^{\text{ext}}_k(\vec{x})}
\mathcal{D}U e^{-S_W} \, \det M \,.
\end{eqnarray}
The functional integration is performed over the lattice
links, but constraining the spatial links belonging to a given time
slice (say $x_t=0$) to be
\be \label{coldwall} U_k(\vec{x},x_t=0) = U^{\text{ext}}_k(\vec{x})
\,,\,\,\,\,\, (k=1,2,3) \,\,, \ee
$U^{\text{ext}}_k(\vec{x})$ being the elementary parallel transports
corresponding to the external continuum gauge potential.
In the previous equations $\vec{A}^{\text{ext}}(\vec{x})$ is the continuum gauge potential of
the external static background field, $S_W$ the standard pure gauge Wilson
action, $S_F$ is the fermion action and $M$ is the fermionic matrix.
The spatial links  are  constrained to the values corresponding to
the lattice version of the external background field, whereas the fermionic fields are not
constrained.
In the case of a static background field which does not vanish at
infinity we must also impose that, for each time slice $x_t \ne 0$,
spatial links exiting from sites belonging to the spatial boundaries
are fixed according to eq.~(\ref{coldwall}). In the continuum this
last condition amounts to the requirement that fluctuations over the
background field vanish at infinity.

We  compute by lattice simulations the derivative
$F^\prime$ with respect to the inverse gauge coupling
\begin{equation}
\label{deriv} F^\prime(\beta) = \frac{\partial
{\mathcal{F}}(\beta)}{\partial \beta} = \left \langle
 \sum_{x,\mu < \nu}
\frac{1}{3} \,  \text{Re}\, {\text{Tr}}\, U_{\mu\nu}(x) \right\rangle_0  \\
  - \left\langle  \sum_{x,\mu< \nu} \frac{1}{3} \,  \text{Re} \, {\text{Tr}} \, U_{\mu\nu}(x)
\right\rangle_{\vec{A}^{\text{ext}}} \,,
\end{equation}
where the subscripts on the averages indicate the value of the
external field. Only unconstrained plaquette are taken into account
in the sum in eq.~(\ref{deriv}).

In the present work we have considered a static constant abelian chromomagnetic field.
The continuum  gauge potential giving rise to a
static constant abelian chromomagnetic field directed along spatial
direction $\hat{3}$ and direction $\tilde{a}$ in the color space can
be written in the following form:
\be \label{su3pot} \vec{A}^{\text{ext}}_a(\vec{x}) =
\vec{A}^{\text{ext}}(\vec{x}) \delta_{a,\tilde{a}} \,, \quad
A^{\text{ext}}_k(\vec{x}) =  \delta_{k,2} x_1 H \,. \ee
In SU(3) lattice gauge theory the constrained lattice links (see
eq.~(\ref{coldwall})) corresponding to the continuum gauge potential
eq.~(\ref{su3pot}) are (choosing $\tilde{a}=3$, i.e. abelian
chromomagnetic field along direction $\hat{3}$ in color space)
\be \label{t3links}
\begin{split}
& U^{\text{ext}}_1(\vec{x}) = U^{\text{ext}}_3(\vec{x}) =
{\mathbf{1}} \,,
\\
& U^{\text{ext}}_2(\vec{x}) =
\begin{bmatrix}
\exp(i \frac {a g H x_1} {2})  & 0 & 0 \\ 0 &  \exp(- i \frac {a g H
x_1} {2}) & 0
\\ 0 & 0 & 1
\end{bmatrix}
\,.
\end{split}
\ee
Since our lattice has the topology of a torus, the magnetic field
turns out to be quantized
\be \label{quant} a^2 \frac{g H}{2} = \frac{2 \pi}{L_1}
n_{\text{ext}} \,, \qquad  n_{\text{ext}}\,\,\,{\text{integer}}\,.
\ee
In the following $n_{\text{ext}}$ will be used to parameterize the
external field strength.

\section{Deconfinement temperature and critical field strength}
\label{deconfinement}

Numerical simulations have been performed on   $32^3\times8$ and
$64\times32^2\times8$ lattices, using computer facilities at the
INFN apeNEXT computing center in Rome.
A slight modified version of the standard
HMC R-algorithm~\cite{Gottlieb:1987mq} has been adopted to
simulate QCD with two degenerate flavors of
staggered fermions of mass $a m_q = 0.075$.
The critical gauge coupling has been determined from the position of
the peak in
$F^\prime[\vec{A}^{\text{ext}}]$ (eq.~(\ref{deriv})), the derivative
of the free energy with respect to the gauge coupling $\beta$, as a
function of $\beta$.
In figure 1
%
%
% FIGURE 1
\FIGURE[ht]{\label{Fig1}
\includegraphics[width=0.85\textwidth,clip]{figure_1.eps}
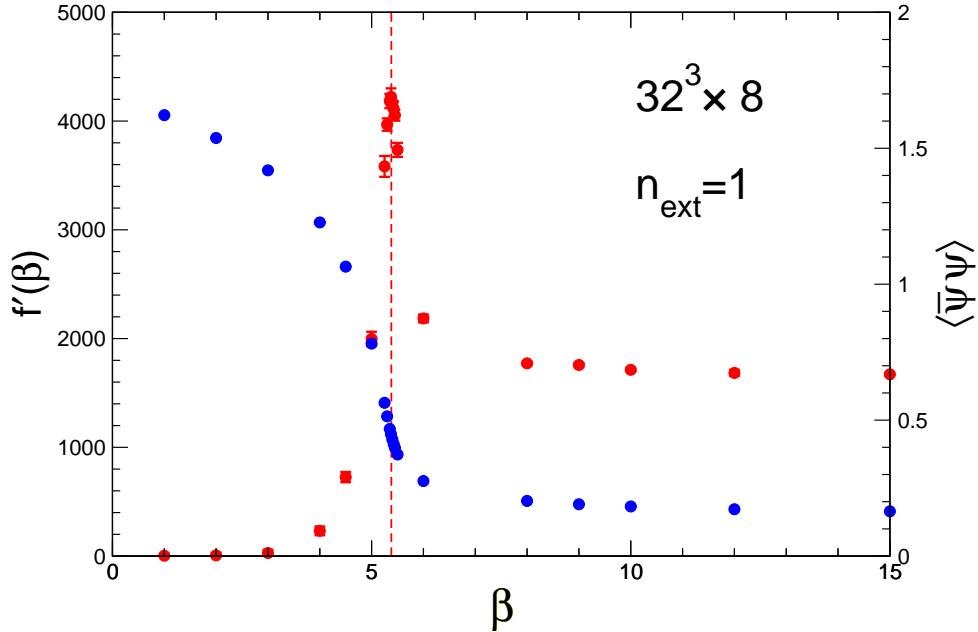
\caption{The derivative of the free energy eq.~(\ref{deriv}) with
respect to the gauge coupling (left axis, blue circles), and the
chiral condensate (right axis, red squares)
versus $\beta$. The vertical line represents the position of the
peak in the derivative of the free energy.} }
we show an example of $F^\prime$ measured for $n_{\text{ext}} = 1$
on a $32^3 \times 8$ lattice. In the same figure we display also the
chiral condensate $\langle \bar{\psi} \psi \rangle$.
We can see that the peak in
the derivative of the free energy corresponds to the drop in the
chiral condensate, the latter being a signal of the transition leading
to chiral symmetry restoration.
The determination of the critical coupling from the peak of the derivative of the free energy
is also consistent with the determinations obtained by studying the Polyakov and the plaquette
susceptibility.
Therefore we may conclude that the
critical coupling of the phase transition can be located by looking
at the peak of the derivative of the free energy and, as in
the case of zero external field and within statistical
uncertainties, a single transition seems to be present
where both deconfinement and chiral symmetry restoration take place.

We have then varied the strength of the external field by
tuning up the parameter $n_{\text{ext}}$ and   searched for
the phase transition signalled by the peak of the derivative of the
free energy finding that the critical coupling shifts
towards lower values  by increasing the external field strength.
On the other hand we have found that an
abelian monopole background field does not have any influence on the critical coupling.
Indeed we have performed numerical simulations in presence of an abelian
monopole background field with monopole charge $n_{\text{mon}}=10$
(again for 2 staggered flavors QCD of mass $a m_q=0.075$). The
critical coupling has been estimated to be $\beta_c = 5.4873(192)$
consistently with the critical coupling $\beta_c = 5.495 (25)$
obtained without any external field.

Once the critical couplings  in correspondence of several
values of the external abelian chromomagnetic field have been obtained
the corresponding critical temperature is given by
\be \label{criticaltemp}
T_c = \frac{1}{a(\beta_c, m_q) L_t}  \,,
\ee
where $L_t$ is the lattice temporal size and $a(\beta_c, m_q)$
is the lattice spacing at the given critical coupling $\beta_c$.
We have  to
face now the problem of fixing the physical scale. In order to reduce
the systematic effects involved in this procedure we   consider
the ratios
\be \label{ratios} \frac{T_c(gH)}{T_c}  \qquad \text{vs.}
\qquad \frac{\sqrt{gH}}{T_c}
\ee
where $T_c$ is the critical
temperature without external field. The above quantities can be
obtained once the ratio of the lattice spacings at the respective
couplings is known. A rough estimate of this ratio can be inferred
by using the 2-loop scaling function $f(g^2)$
\be
\label{asympscaling}
f(g^2) = (b_0
g^2)^{-b_1/2b_0^2}  \,\, \exp\left(- \frac{1}{2 b_0 g^2} \right)
\ee
with
\be
\label{coeffs}
\begin{split}
&b_0 = \frac{1}{16 \pi^2} \left[ 11 \frac{N_c}{3} - \frac{2}{3} N_f \right] \\
&b_1 = \left(\frac{1}{16 \pi^2} \right)^2 \left[ \frac{34}{3} N_c^2
-
    \left( \frac{10}{3} N_c + \frac{N_c^2 -1}{N_c} \right) N_f \right]
  \, ;
\end{split}
\ee
$N_c$ is the number of colors and $N_f$ is the number of
flavors.
A better estimate could be
obtained by exploiting an improved scaling
function $f(g^2) (1+c_2 \hat{a}(g)^2 + c_4 \hat{a}(g)^4)$. We do not
know, however, the values of $c_2$ and $c_4$ for $N_f = 2$. In a
first approximation we will fix $c_2$ and $c_4$ to their quenched
values~\cite{Allton:1996kr,Edwards:1998xf}.

In figure~\ref{Fig7}
%
%
%
% FIGURE 7
\FIGURE[ht]{\label{Fig7}
\includegraphics[width=0.85\textwidth,clip]{figure_7.eps}
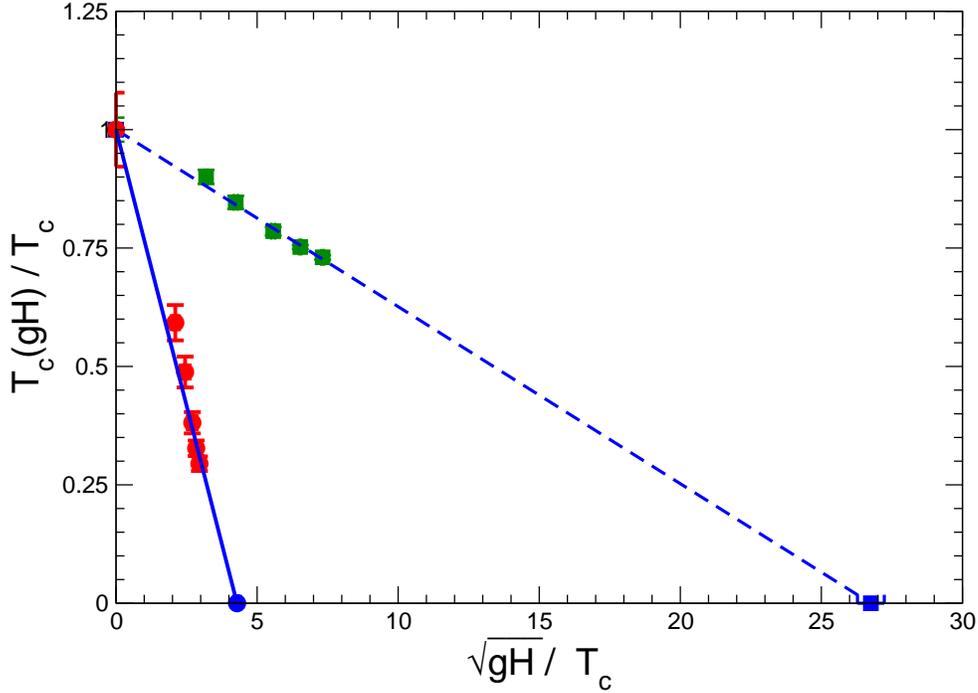
\caption{The critical temperature $T_c(gH)$ at a given strength of
the chromomagnetic background field in units of the critical
temperature $T_c$ without external field versus the square root of
the strength of the background field in the same units. Red circles
are obtained by adopting the improved scaling function. The blue
line is the linear best fit. The blue circle  on the horizontal axis
is the linear extrapolated values for the critical background field.
Green squares are obtained by adopting the 2-loop scaling function.
The blue dashed line is the linear best fit. The blue square  on the
horizontal axis  is the linear extrapolated values for the critical
field.} }
the quantities reported in eq.~(\ref{ratios}) are displayed for both
choices described above, i.e. 2-loop asymptotic scaling and improved
scaling.
Our  main result  is that,  even in presence of
dynamical quarks, the critical temperature decreases  with the
strength of the chromomagnetic field. Furthermore a linear fit to our
data can be extrapolated to very low temperatures leading to the
prediction of a critical field strength above which strongly
interacting matter should be deconfined at all temperatures.
Nevertheless, as can be appreciated from figure~\ref{Fig7}, the exact
value of the critical field strength is largely dependent on the
choice of the physical scale:  assuming the 2-loop
scaling function we obtain $\sqrt{gH_c}/T_c  = 26.8 (5)$,
while  using the
improved scaling function we obtain $\sqrt{gH_c}/T_c  = = 4.29 (10) $;
if the deconfinement
temperature at zero field strength is taken to be of the order of
$170$ MeV, that means $\sqrt{gH_c}$ in the range $0.7 - 4.5$ GeV.
Hence in order to get a reliable estimate of the critical
field strength requires to
get a more reliable estimate of the physical scale of our lattices.

To conclude this section we consider our measurements of the chiral
condensate. In figure~\ref{Fig8}
%
%
% FIGURE 8
\FIGURE[ht]{\label{Fig8}
\includegraphics[width=0.85\textwidth,clip]{figure_8.eps}
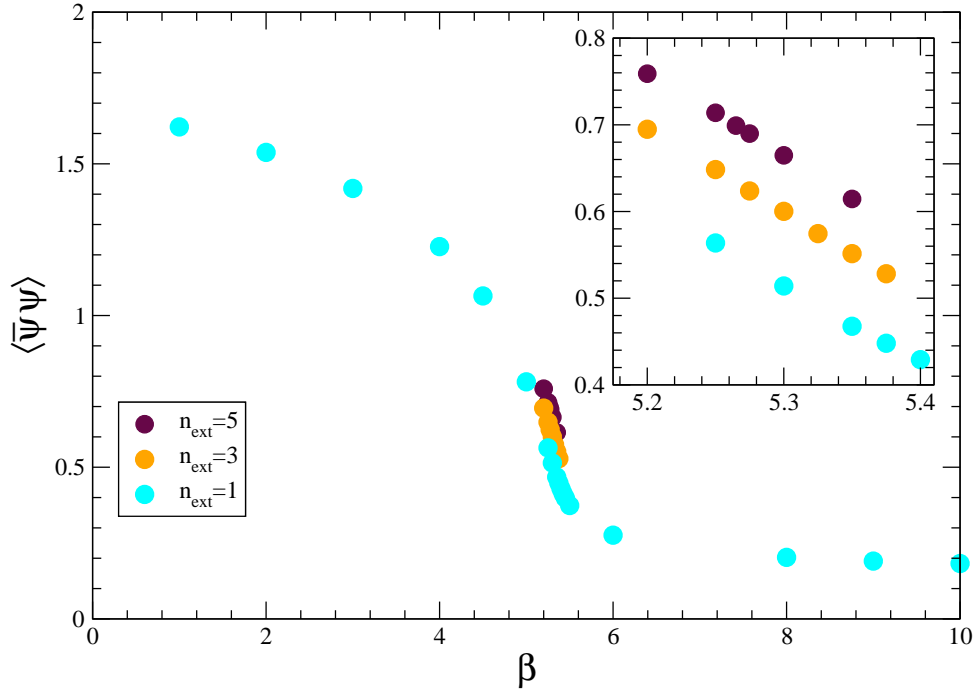
\caption{The chiral condensate  versus $\beta$
in correspondence of some values of the constant chromomagnetic
background field. In the inset the region corresponding to the phase
transition has been magnified.} }
we display the chiral condensate versus the gauge coupling in
correspondence of some values of the external field strength.
Numerical data show that, at least in the critical region, the value
of the chiral condensate depends on the strength of the applied
field. Similar results have been found in ref.~\cite{Alexandre:2001pa}.

\section{Summary and Conclusions}
We have studied how a constant chromomagnetic field
influences the QCD dynamics. By focusing on the
finite temperature theory  we have found that, analogously to what
happens in the pure gauge theory~\cite{Cea:2005td}, the critical
temperature depends on the strength of the constant chromomagnetic
background field. More specifically, the critical
temperature  decreases as the external field is increased
and, as an extrapolation of our results, for strong enough field
strengths the system is always deconfined.
A rough estimate of this critical field strength turns out  to be of the
order of 1~GeV, which is a typical QCD scale~\cite{Kabat:2002er}.

What is more, by comparing the critical couplings determined from the derivative
of the free energy functional with those determined from the
susceptibility of the chiral condensate and of the Polyakov loop we
have ascertained that, even in presence of an external
chromomagnetic background field and at least up to the field
strengths explored in the present work, the critical temperatures
where deconfinement and chiral symmetry restoration take place
coincide within errors.

Another intriguing aspect that deserves further studies  is the dependence  of the
chiral condensate on the chromomagnetic field strength.

%\bibliography{qcd}
\providecommand{\href}[2]{#2}\begingroup\raggedright\endgroup

\end{document}